\documentclass[aps,epsf]{revtex4}
\usepackage{psfig}

\begin{document}

\title{On J. Goodman's comment to \\ {\em Language Trees and Zipping}}
\author{D. Benedetto$^1$, E. Caglioti$^1$ and V. Loreto$^2$}
\date{\today}

\affiliation{$^1$ ``La Sapienza'' University, Mathematics Department ,
P.le A. Moro 5, 00185 Rome, Italy\\$^2$ ``La Sapienza'' University,
Physics Department, P.le A. Moro 5, 00185 Rome, Italy and INFM, Center
for Statistical Mechanics and Complexity, Rome, Italy}

\vspace{-5pt}

\begin{abstract}
Motivated by the recent submission to cond-mat archives by J. Goodman
whose results apparently discredit the approach we have proposed in a
recent paper ({\em Phys. Rev. Lett.}  {\bf 88}, 048702 (2002)), we
report the results of the same experiment performed by Goodman using
three different data compression schemes. As a matter of fact the
three zippers display the same efficiency Goodman obtained using Naive
Bayesian Methods and not, as Goodman claimed, an efficiency three
times smaller. We point out the question of the extreme generality of
approaches based on data compression techniques and we list a large
range of potential applications, including those of interest for the
physics community.
\end{abstract}
\maketitle
\section{Introduction}

In this short note we reply to J. Goodman's comments on a paper we
have recently published (``Language Trees and Zipping''~\cite{PRL}).

First of all we wish to apologize with the users of electronic
archives for an use of them that we judge improper. Electronic
archives should be used to spread in a fast way scientific knowledge
and surely they are not intended as a large audience where raising
polemics whose scientific content is at least doubtful.  We shall
accordingly try to keep the discussion on a scientific basis which
could add some technical information for the reader genuinely
interested in knowing whether our method could be used and for which
purposes. Nevertheless we shall just point out some inconsistencies in
the Goodman's rationale.

In a recent cond-mat submission~\cite{JG-cond-mat} (anticipated by an
{\em Open letter to the editors of Physical Review Letters} that we
report in Appendix) J. Goodman strongly criticizes our paper on
the basis of three main points:

\begin{itemize}

\item[$\bullet$] The paper does not relates to physics and it should
not be published; 

\item[$\bullet$] Goodman reviews the literature to conclude that ideas
like the ones discussed in our paper are generically not novel;

\item[$\bullet$] Goodman provides experimental results which
apparently discredit our approach and our methods.

\end{itemize}

In this Letter we shall mainly discuss in detail the last point raised
by Goodman, i.e. the technical point and probably the only one
deserving attention from a scientific community.  As for the other two
points Goodman raises let us just notice the following.

The question whether a paper {\em unrelated} to physics should or
should not be published on a journal like Physical Review Letters is
beyond our judging possibilities. Probably the criticism should be
addressed more to the Editors of Physical Review Letters (see the Open
Letter to them in Appendix). Nevertheless Physical Review Letters has
a section named {\em Interdisciplinary Physics: Biological Physics,
Quantum Information, etc.} where we naively thought our paper could
fit. If then one considers that publication on Physical Review Letters
is subject to a tough peer review probably we could not be blamed for
having attempted this submission.

Goodman says:

{\em Physics journals should not be publishing articles that have
nothing to do with physics.  Of course, it is completely reasonable to
publish applications of physics to other fields (both because this
alerts other physicists to the possibility of applying their knowledge
and because those in the field of interest may have difficulty
understanding the terminology or techniques).  It is also completely
reasonable to publish the use of non-physics techniques applied to
physics in a physics journal.  But this paper applies computer science
techniques (gzip!) to computer science problems.  This seems extremely
inappropriate. One might argue that the paper discusses entropy, a
concept taken from physics.  But the concept was taken from physics 50
years ago, at the dawn of computer science, and there is nothing
physics-specific in the use of entropy in this paper; indeed, the use
is entirely in the information theory/language modeling/computer
science meaning of the word.}

Except from learning that the concept of entropy now belongs to
Computer Science (sic !), there is one important point to stress: our
method provides a general framework to analyze for instance a
generic time-series: could one imagine a more physical application
than the study of predictability in dynamical systems or any other
phenomenon, e.g. experimental investigations, where one typically has
access to the system only through a measuring device which produces a
time record of a certain observable, i.e. a sequence of data?

Apart from this there is something obscure in Goodman's comment.
There are typically several reasons to criticize a paper: (a) the
paper is wrong: there are evident mistakes; (b) the paper is not
publishable because the results it reports have already been obtained
and published somewhere else (and in this case one reports the
relevant references). If the paper is correct but its relevance is
limited one typically does not waste his time trying to put in a black
light something that the community will anyway ignore.

From this point of view it is not clear what disturbs Goodman.
Apparently the paper is correct from his point of view. Apparently he
is not able to quote relevant references where a similar approach has
been proposed. He is only able to say:

\vspace{0.2cm} ``The only idea in this paper that is not very well
known in the field is the idea of actually using a standard
compression tool to do the classification.  Still, even this idea
dates back (at least) to 1995, when Ken Lang and Rich Caruana tried
out the idea for doing newsgroup classification.  In the end though,
they ended up using Naive Bayes classifiers~\cite{Lang}.  {\em They
didn't bother to publish the compression idea because they thought it
was better viewed as an interesting thought than as a serious
classification method}. Still, the idea of using compress got around a
bit: see an introductory tutorial by a well known practitioner in this
area, Tom Mitchell~\cite{Mitchell}.  Admittedly, however, this
technique is not that widely known, because computer scientists don't
typically try anything that crude -- in a couple of hours (or less),
we can build from scratch tools that work better than this.''
(we have only added the italic style).

\vspace{0.2cm}

Let us now come to the technical part of the paper discussing the last
point raised by Goodman, namely the experimental results that
apparently discredit our method. Here Goodman's claim is as follows:
``If the Benedetto, Caglioti, Loreto method fails miserably against
the simplest method computer scientists could think of, then why all
this noise about it?''.

\section{Results on Classification by subject}
 
In this section we report the results of the same {\em experiments}
performed by J. Goodman using our method for the analysis of the
newsgroup messages available from {\tt
http://www.ai.mit.edu/{\linebreak[0]}\~{}jrennie/{\linebreak[0]}20Newsgroups}.
(Exactly the same version used by Goodman containing 18828 messages).

It is important to stress that Goodman is criticizing our paper in a
very odd way. He claims that our technique fails on an experiment (the
subject classification) that was not described in our paper. In his
position we would have criticized the paper on its content and not on
something completely different. We are nevertheless able to perform
subject classification but we have never published how the technique
works in this case. It is then quite natural for us to wonder what
precisely Goodman did when he says: {\em ``For zipping, I applied the
algorithm of~\cite{PRL}}''.

On the other hand, supposing Goodman used the technique described in
the paper for authorship attribution, it is hard for us to understand
how it is possible to make our method working so badly.  We have
nevertheless accepted the challenge and analyzed the same newsgroup
messages as Goodman.

The technique we use (independently of the zipper considered) is as
follows:

\begin{itemize}

\item[$\bullet$] For each unknown texts, consider the collection of
texts obtained attaching it to each one of the texts considered as
reference;

\item[$\bullet$] Zip the resulting files as well as all the reference
files separately and compute, for each reference file the difference
between the size (in bytes) of the new total file zipped and the
reference file zipped;

\item[$\bullet$] Write for each unknown text the ranking of all the
reference texts. Once one has the entire ranking for each unknown text
one can start the attribution of each unknown text to the most
pertinent category.  Several paths are of course possible in agreement
that with the fact that any classification has some level of
arbitrary.  The perfect classification does not exist: the proof lies
in the fact that one can always refine the list of categories ending
up with one specific category for each specific item in the list.

\end{itemize}

We use the criterion to associate the unknown text to the category of
the reference text that was ranked as first in its particular ranking
(as described for authorship attribution in our paper).  We have
performed the analysis using several zippers in order to compare the
results and show the robustness of the approach.  We have used in
particular the following data compression schemes: gzip, a modified
version of gzip (denoted as bcl) which differs from gzip for the fact
of simply reading the second file instead of zipping it and on the
Huffman optimization~\cite{long}. We have also used the method
proposed by Merhav and Ziv~\cite{merhav-ziv} to measure the relative
entropy between two sequences of characters.  The results are shown in
Table I for a sample of $200$ newsgroup messages chosen in a random
way.

\begin{table}[tb]
\begin{tabular}{|l|cccc|cccc|cccc|}

\hline
{\centering {\sf  }}  
&{\centering {\sf  }}
&{\centering {\sf  }}  
&{\centering {\sf  {\large gzip}}} 
&{\centering {\sf  }} 
&{\centering {\sf  }} 
&{\centering {\sf  }} 
&{\centering {\sf {\large bcl}}} 
&{\centering {\sf  }}
&{\centering {\sf  }}
&{\centering {\sf  }}  
&{\centering {\sf {\large mz}}}
&{\centering {\sf  }} 
\\\hline
{\centering {\sf N. of messages}} 
&{\centering {\sf  Succ.{\bf 1}}}
&{\centering {\sf  \%}}  
&{\centering {\sf  Succ.{\bf 2}}}
&{\centering {\sf  \%}} 
&{\centering {\sf  Succ.{\bf 1}}}
&{\centering {\sf  \%}}  
&{\centering {\sf  Succ.{\bf 2}}}
&{\centering {\sf  \%}} 
&{\centering {\sf  Succ.{\bf 1}}}
&{\centering {\sf  \%}}  
&{\centering {\sf  Succ.{\bf 2}}}
&{\centering {\sf  \%}} 
\\\hline
{\centering {\sf  200}} 
&{\centering {\sf  173}}
&{\centering {\bf  86.5\%}}  
&{\centering {\sf  181}} 
&{\centering {\bf  90.5\% }} 
&{\centering {\sf  173}}
&{\centering {\sf  86.5\%}}  
&{\centering {\sf  180}} 
&{\centering {\sf  90.0\% }} 
&{\centering {\sf  174}}
&{\centering {\sf  87.0\%}}  
&{\centering {\sf  183}} 
&{\centering {\sf  91.5\%}} 
\\\hline
\end{tabular}
\label{ric-autore}
\caption{{\bf Newsgroup classification}. For each zipper we
report the number of different texts considered and two measures of
success. Number of success $1$ and $2$ are the numbers of times
another message in the same newsgroup was ranked in the first position
or in one of the first two positions respectively.}
\end{table}

As a matter of fact, using only the information coming from the
document ranked in the first position (and thus not exploiting the
entire information of the ranking list) the three zippers display the
same efficiency Goodman obtained using Naive Bayesian Methods and not,
as Goodman claimed, an efficiency three times smaller.  As for the
computational time we agree with Goodman that data compression
techniques are slower than other procedures but here we are interested
in testing the efficiency of the algorithms for a large variety of
problems (not only newsgroup classification).  We think that the
question of speed of the algorithms should be addressed once one has
identified the best algorithm (from the point of view of the
efficiency) for a large set of problems. It would interesting for
instance to know how Naive Bayesian methods perform in the set of
problems we have considered in~\cite{PRL}. For instance in the case of
language classification, having only one short text available per
language, it is not clear how a Naive Bayesian method could sort out a
measure of remoteness between pairs of texts to be used after for a
tree representation.  In this case we can guess that the method would
suffer the lack of long training texts and this could also be true for
the problem of authorship recognition where the number of texts
available per author is very limited.

\section{Conclusions}

Before concluding several remarks are in order.  The results shown
above demonstrate as algorithms based on data compression techniques
are highly competitive.  It should be stressed how Goodman did not
performed the same experiments we report in~\cite{PRL} but he tried to
discredit our approach performing a very focused test on one
particular problem: the newsgroup classification. Unfortunately in
that particular case his claim was not correct and our efficiency
equals that of Naive Bayesian Approaches for this specific problem. It
is worth to stress how on much more controlled and clean corpora the
efficiency of our method raises considerably. We refer the reader to a
long paper~\cite{long} where we discuss in detail all the issues
simply sketched in~\cite{PRL} (as well as the classification by subject
problem applied to large law corpora) and we compare the results
obtained with different data compression schemes: gzip, bzip2,
compress, merhav-ziv, etc.

From our point of view the situation could be interpreted as follows:
it is not correct to neglect some particular technique on the only
basis that it is the simplest technique a computer scientist can think
of.  Our point of view is that the use of data compression techniques
represents a very general tool in several areas and not only in the
narrow field of newsgroup classification.  Always remaining in the
field of computational linguistics there have been several
contributions showing how data compression techniques could be useful
in solving different problems (an incomplete list would include
~\cite{bell,teahan,juola,el-yaniv,Kontoyiannis,nevill-manning,thaper}):
language recognition, authorship recognition, language classification,
classification of large corpora by subject, etc. In this spirit our
paper~\cite{PRL} represents only one possible example which could
hopefully give a contribution to this challenging field.

It is also important to stress how data compression oriented
techniques only represent one sector in the large area of language
modeling for the solution of linguistic motivated problems. Also the
methodologies using "entropic" or statistical concepts are very many
and used in very different contexts: n-grams, hidden Markov modeling,
Bayesian approaches, context-free grammars, neural networks just to
quote some existing approaches. We refer to~\cite{rosenfeld} for a
recent overview about statistical language modeling.

Of course the possibilities of data-compression based methods go
beyond computational linguistics. More in general since these methods
apply to generic sequences of characters, we are not limited to
consider only texts (and linguistic applications) but we can consider
other cases where the strings of characters represent a different kind
of coding: time sequences, genetic sequences (DNA or proteins) etc.
These features are potentially very important for fields where the
human intuition can fail: DNA and protein sequences, geological time
series, stock market data, medical monitoring, etc.  In summary the
fields of application range from time-series analysis (we refer the
reader to~\cite{yak} for an application to time-series generated by
some specific dynamical systems), to theory of dynamical
systems~\cite{benci}, genetic problems (here also an incomplete list
would include~\cite{li,grumbach,Loewenstern}). One of the advantages
of using a data compression scheme is that in this case one can face
all the problems mentioned above in the same framework.

\newpage

\appendix{Appendix I}

\vspace{1.5cm}

\centerline{An open letter to the editors of Physical Review Letters}

\vspace{0.5cm}
(The letter appeared only for a few days on the site: {\tt
http://research.microsoft.com/\~{}joshuago/} and then mysteriously
disappeared, substituted by a more serious comment sent to the Editors
of Physical Review Letters as well as to cond-mat archives.)
\vspace{0.5cm}

\begin{verbatim}

Date: Tue, 12 Feb 2002 15:17:12 -0800
From: Joshua Goodman <joshuago@microsoft.com>
To: benedetto@mat.uniroma1.it, caglioti@mat.uniroma1.it, loreto@roma1.infn.it
Subject: Open Letter to the editors of Physical Review Letters


An open letter to the editors of Physical Review Letters

 

Dears Sirs,

 

I wish to commend you on your open-minded interdisciplinary approach

to science.  In particular, I was extremely impressed that you

published an article on Computational Linguistics and Machine Learning

in a journal ostensibly devoted to physics ("Language Trees and

Zipping" by Dario Benedetto, Emanuele Caglioti, and Vittorio Loreto,

28 January 2002, (available at

http://babbage.sissa.it/abs/cond-mat/0108530) with the following

Abstract:

 

  In this Letter we present a very general method for extracting

  information from a generic string of characters, e.g., a text, a DNA

  sequence, or a time series. Based on data-compression techniques, its

  key point is the computation of a suitable measure of the remoteness

  of two bodies of knowledge. We present the implementation of the

  method to linguistic motivated problems, featuring highly accurate

  results for language recognition, authorship attribution, and language

  classification.

 

I myself attempted to publish the following letter on Physics in

several journals, including Computational Linguistics, the Machine

Learning Journal, and Computer Speech and Language:

 

"Measurement of Gravitational Constant Using Household Items" by

Joshua Goodman, unpublished manuscript.

 

  In this Letter we present a method for measuring the gravitational

  constant ("g") using household items, such as old shoes and a

  microwave oven.  Using the timer on the microwave oven, we were able

  to measure the time for the old shoe to reach the ground from several

  different heights.  We were able to determine that "g" equals 34 feet

  per second per second, plus or minus 3.  This is within 10% of the

  current state of the art in physics.  We noticed that our curves were

  not quite parabolic, and we speculate that this is evidence of a

  wormhole in the region of the experiment.

 

One of the reviews I received was very positive ("I commend you on

this excellent research.  Some might argue that it is irrelevant to a

journal such as ours, but physics, and 'g' in particular, effect us

each and every day, for instance, as we walk or pour coffee.

Furthermore, since almost all readers of this journal have had a high

school (or even college) physics class, and all of us have and use

household items such as these, it is sure to be of interest to the

readership.") Unfortunately, the other reviewers were far more

critical.  One reviewer wrote "While it is clear that physicists

should be writing papers about natural language processing and

statistical modeling, this is because they are so much smarter than we

computer scientists are.  The reverse is obviously not true."  Another

reviewer was even too lazy to read the paper, writing "As everyone

knows, it takes dozens (or even hundreds) of authors to do any good

work in physics.  I find it difficult to believe that a single

authored letter is worth reading."  And another wrote "I remember from

my coursework something about Newton's Method and Gradient Descent.

How come neither of these are cited in this paper?"  With comments

such as those, it seems that reviewers in our field are not even

competent to review physics papers, while clearly yours know enough to

review computer science papers.

 

Obviously, those in my field are not nearly as broad minded or well

rounded as those in yours.  I commend you on your openness, and hope

that you will soon expand to cover other areas of interest, such as

economics, psychology, and even literature.

 

Sincerely yours,

 

Joshua Goodman

 

----------------------------------------------------------------------


\end{verbatim}

\end{document}